\documentstyle[12pt,aaspp4,epsfig]{article}

\newcommand{\etal}{{\it{et al.}}~}

\newcommand{\hst}{{\it{HST }}}

\def\gtsima{$\, \buildrel > \over \sim \,$}
\def\ltsima{$\, \buildrel < \over \sim \,$}
\def\simgt{\lower.5ex\hbox{\gtsima}}
\def\simlt{\lower.5ex\hbox{\ltsima}}

\begin{document}

\title{Stellar Populations in Three Outer
Fields of the LMC \footnote{Based
on observations with the NASA/ESA {\it Hubble Space Telescope},
obtained at
the Space Telescope Science Institute, operated by AURA Inc under
contract
to NASA}}
\author{
Marla~C.~Geha\altaffilmark{2},
Jon~A.~Holtzman\altaffilmark{2},
Jeremy~R.~Mould\altaffilmark{3},
John~S.~Gallagher~III\altaffilmark{4},\\
Alan~M.~Watson\altaffilmark{5},
Andrew~A.~Cole\altaffilmark{4}
Carl~J.~Grillmair\altaffilmark{6},
Karl~R.~Stapelfeldt\altaffilmark{6},
Gilda~E.~Ballester\altaffilmark{7},
Christopher~J.~Burrows\altaffilmark{8},\\
John~T.~Clarke\altaffilmark{7},
David~Crisp\altaffilmark{6},
Robin~W.~Evans\altaffilmark{6},
Richard~E.~Griffiths\altaffilmark{9},
J.~Jeff~Hester\altaffilmark{10},\\
John~G.~Hoessel\altaffilmark{4},
Paul~A.~Scowen\altaffilmark{10},
John~T.~Trauger\altaffilmark{6},
and
James~A.~Westphal\altaffilmark{11}
}
 
\altaffiltext{2}{Department of Astronomy, New Mexico State
University, Dept 4500
 Box 30001, Las Cruces, NM 88003, mgeha@nmsu.edu, holtz@nmsu.edu}
\altaffiltext{3}{Mount Stromlo and Siding Spring Observatories,
Australian National University, Private Bag, Weston Creek Post
Office, ACT 2611, Australia,
jrm@mso.anu.edu.au}
\altaffiltext{4}{Department of Astronomy, University of Wisconsin
-- Madison, 475 N. Charter St., Madison, WI 53706,
jsg@tiger.astro.wisc.edu,cole@ninkasi.astro.wisc.edu,hoessel@jth.
astro.wisc.edu}
\altaffiltext{5}{Instituto de Astronom\'\i a UNAM, J. J. Tablada
1006, Col. Lomas de Santa Maria, 58090 Morelia, Michoac\'an,
Mexico, alan@astrosmo.unam.mx}
\altaffiltext{6}{Jet Propulsion Laboratory, 4800 Oak Grove Drive,
Pasadena,
CA 91109, carl@wfpc2-mail.jpl.nasa.gov, dc@crispy.jpl.nasa.gov,
rwe@wfpc2-mail.jpl.nasa.gov, krs@wfpc2-mail.jpl.nasa.gov,
jtt@wfpc2-mail.jpl.nasa.gov}
\altaffiltext{7}{Department of Atmospheric, Oceanic, and Space
Sciences,
University of Michigan, 2455 Hayward, Ann Arbor, MI 48109,
gilda@sunshine.sprl.umich.edu, clarke@sunshine.sprl.umich.edu}
\altaffiltext{8}{Astrophysics Division, Space Science Department,
ESA \& Space
Telescope Science Institute, 3700 San Martin Drive, Baltimore, MD
21218,
burrows@stsci.edu}
\altaffiltext{9}{Department of Physics, Carnegie Mellon
University, 5000 Forbes
Ave, Pittsburgh, PA 15213}
\altaffiltext{10}{Department of Physics and Astronomy, Arizona
State University,
Tyler Mall, Tempe, AZ 85287, jjh@cosmos.la.asu.edu,
scowen@tycho.la.asu.edu}
\altaffiltext{11}{Division of Geological and Planetary Sciences,
California Institute of Technology, Pasadena, CA 91125,
jaw@sol1.gps.caltech.edu}

\newpage
\begin{abstract}
We present \hst photometry for three fields in the outer disk of the
LMC extending approximately four magnitudes below the faintest main
sequence turnoff.  We cannot detect any strongly significant
differences in the stellar populations of the three fields based on
the morphologies of the color-magnitude diagrams, the luminosity
functions, and the relative numbers of stars in different evolutionary
stages.  Our observations therefore suggest similar star formation
histories in these regions, although some variations are certainly
allowed.  The fields are located in two regions of the LMC: one is in
the north-east field and two are located in the north-west.  Under
the assumption of a common star formation history, we combine the
three fields with ground-based data at the same location as one of the
fields to improve statistics for the brightest stars. We compare this
stellar population with those predicted from several simple star
formation histories suggested in the literature, using a combination
of the R-method of Bertelli \etal (1992) and comparisons with the
observed luminosity function. The only model which we consider that is
not rejected by the observations is one in which the star formation
rate is roughly constant for most of the LMC's history and then
increases by a factor of three about 2 Gyr ago. Such a model has
roughly equal numbers of stars older and younger than 4 Gyr, and thus
is not dominated by young stars. This star formation history, combined
with a closed box chemical evolution model, is consistent with
observations that the metallicity of the LMC has doubled in the past 2
Gyr.
\end{abstract}

\section{INTRODUCTION}
The star formation history of field stars in the Large Magellanic
Cloud (LMC) contains much information about the formation and dynamics
of our closest galactic neighbor.  Studies by Stryker
\markcite{stry84}(1984), Bertelli \etal \markcite{bmcb92}(1992),
Westerlund \etal \markcite{wll95}(1995), and Vallenari \etal
\markcite{vcbo96a}\markcite{vcba96b}(1996a,b), among others, conclude
that the LMC field contains a majority of young to intermediate age
stars overlying a minority old population.  Bertelli \etal
\markcite{bmcb92}(1992) favor a star formation history in which the
star formation rate, initially at a constant low level, increases by a
factor of ten in the last few billion years.  This star formation
history produces a stellar population reminiscent of the bimodal age
distribution of LMC globular clusters (van den Bergh
\markcite{vand91}1991; Girardi \etal \markcite{gira95}1995).
Vallenari \etal \markcite{vcba96b}(1996b) find tentative evidence that
the onset of this increase in star formation is correlated with
position in the LMC, and suggest that this correlation might arise if
star formation is triggered by tidal interactions with the Small
Magellanic Cloud.  Elson \etal \markcite{egs97}(1997) present \hst
observations of a field in the bar of the LMC and find evidence for an
additional younger population of stars which is not observed in the
outer regions of the LMC field.  A clearer understanding of stellar
populations throughout the LMC should provide clues about the age and
formation history of the LMC as well as about the mechanisms which
trigger star formation in this galaxy.

We have observed three fields in the LMC with the Wide
Field Planetary Camera 2 on the {\it Hubble
Space
Telescope} to determine whether these regions share a similar
formation history.  The fields are all located in the outer
regions
of the LMC at roughly the same radial distance from the LMC bar.
These
observations extend several magnitudes below the main sequence
turnoff and provide a significant improvement over ground-based
studies.

Gallagher \etal \markcite{gall96}(1996)
present the color-magnitude diagram for one of these fields,
and find that the width of the upper main sequence is consistent
with a
star formation rate which is roughly constant for the last few
billion
years.  In
addition, they suggest that a small burst of star formation
occurred 2 Gyr
ago, leading to a distinct subgiant branch seen one
magnitude brighter than the faintest main sequence turnoff.  The
lack of evidence in the \hst data for a strong star
formation burst is in apparent contrast to previously determined
star formation histories and results which indicate the
metallicity of the LMC has nearly doubled in the past 2 Gyr
(Dopita \etal \markcite{dopi97}1997).

Holtzman \etal \markcite{holt97}(1997) analyze the luminosity
function of the same \hst field to constrain its initial mass
function and star formation
history.  By comparing the luminosity function
of the lower main
sequence to stellar models, they constrain the IMF
slope, $\alpha$ ($dN/dM \propto M^\alpha$), to $-3.1 \leq \alpha
\leq
-1.6$ in the mass range $0.6 \leq M \leq 3M_{\odot}$.  Assuming a
Salpeter IMF ($\alpha= -2.35$), they derive a star formation
history from the
entire observed luminosity function.  They favor a star formation
history 
in which the star formation rate is roughly constant
for 10
Gyr
and then increases by a factor of three for the past 2 Gyr,
resulting in a stellar population with comparable numbers of
stars older
and younger than 4 Gyr.  This is in contrast to Bertelli
\etal \markcite{bmcb92}(1992), whose preferred star formation
history produces a
predominantly young ($\leq 4$ Gyr)  stellar population.  
Holtzman \etal find that a predominantly young population fits
the
\hst
observations only if the IMF
slope is steeper, with $\alpha \simlt -2.75$.

In $\S 2$, we present \hst observations of the three LMC fields.
In $\S 3$, we quantitatively compare the
stellar
populations in
these fields and show that they are statistically
indistinguishable.  In $\S 4$, we
compare our
observations with several possible star formation histories, 
using the R-method of
Bertelli \etal \markcite{bmcb92}(1992) in combination with 
comparisons between the model and observed luminosity functions. 
Our
derivation  
of the star formation history is an improvement over
previous \hst determined formation histories as we use
ground-based data
to supplement observations at the brightest magnitudes.

\section{OBSERVATIONS}
Observations were made with the Wide Field Planetary Camera 2 of
the {\it Hubble Space Telescope} between May 1994 and December
1995
through the F555W ($\sim V$) and F814W ($\sim I$) filters.  Total
exposure times were
4000s, 2500s, and 1000s in each filter for Fields 1, 2, and 3,
respectively. 
Observations through each filter were split into
three or more separate exposures to allow identification and
removal of cosmic ray events.  In Figure 1, an image of the LMC
shows the approximate positions of the three fields.  A previous
analysis of Field 1 has been presented by Gallagher
\etal \markcite{gall96}(1996) and Holtzman \etal
\markcite{holt97}(1997).

The positions and exposure times for each
field 
are listed in Table 1.  The data
were processed using standard reduction techniques described in
Holtzman \etal \markcite{holt95a}(1995a).  This process includes
a small correction
for analog-to-digital errors, overscan and bias subtraction, dark
subtraction, a small shutter shading correction, and flat
fielding.  In each filter, the images were combined and cosmic
ray events were removed based on the expected variance from
photon
statistics and read noise.

\subsection{Photometry}
A combination of profile-fitting and aperture photometry was chosen to
give good photometry at both bright and faint signal levels.  Since
the F555W and F814W images were roughly equally deep, stars were found
for each field on the summed frame of these two images.  Due to
structure in the point spread function (PSF), objects found in the
area surrounding the peak of bright stars were rejected.  Using this
star list, profile-fitting photometry was performed on each frame.
The model PSFs were the same as those used by Holtzman \etal
\markcite{holt97}(1997).  Profile-fitting results were then used to
subtract all stars from the images.  Final magnitudes were determined
by adding each star individually back into the subtracted frame and
performing aperture photometry with a 2 pixel radius aperture.
Aperture corrections to a 0.5 arcsecond radius aperture were
individually determined for the four WFPC2 chips from bright, isolated
stars.  We estimate the maximum error of this correction to be a few
hundredths of a magnitude.  Instrumental magnitudes were transformed
into V and I magnitudes using the transformations given by Holtzman
\etal \markcite{holt95b}(1995b).

To convert into absolute magnitudes, we adopt a distance modulus of
18.5 derived by Panagia \etal \markcite{pana91}(1991) from SN1987A .
A re-evaluation of the Cepheid distance calibration, using Hipparcos
parallaxes, suggests a slight upward revision in the LMC distance
modulus to $18.57\pm 0.11$ (Madore \& Freedman \markcite{mf97}1997), however, our conclusions are insensitive to errors of this order in
the distance modulus.  Schwering \& Israel \markcite{si91}(1991)
determine a foreground reddening of $E(B-V)=0.07$ towards Field 1 and
variations less than 0.02 in $E(B-V)$ between the three fields.
Allowing for a small amount of internal extinction, we adopt
$E(B-V)=0.1$ with a corresponding extinction of $A_V = 0.31$.

To estimate completeness, a set of artificial stars tests was
performed.  At a series of different brightnesses, artificial stars
were added to each frame in an equally spaced grid and the frames were
run through the photometry routine described above.  The grid spacing
was chosen so that artificial stars did not add significantly to
crowding in the field; 121 stars were placed on the PC and 529 stars
on each of the WFs.  The resulting photometry list was compared to the
input list and the completeness level was determined as described by
Holtzman \etal \markcite{holt97}(1997). We estimate the 90\%
completeness level to be at $m_V\sim 26$ in Fields 1 and 2, and $m_V
\sim 24.5$ in Field 3 due to the lower exposure time.  We restrict our
analysis to stars brighter than these limits.  The fraction of
detected artificial stars and their associated errors were tabulated
as a function of magnitude and are used in $\S 4.2$ to simulate
observed stellar populations.  These errors include systematic errors
due to crowding and random errors from Poisson statistics.
Observational errors for our simulations (discussed below) are determined by
randomly sampling from these error distributions.  The stellar density in
these regions of the LMC is low and the fields are not crowded (see
Holtzman \etal\markcite{holt97}1997 for an image of Field 1).  For stars
brighter than $m_V \leq 18.0$, our observations are not representative
due to saturation and the small WFPC2 field of view.

In $\S 4.2$, we correct for small number statistics at the brightest
magnitudes using ground based data taken with the Mount Stromlo and
Siding Spring Observatories 1m telescope (Stappers \etal
\markcite{stap97}1997).  Three fields adjacent to and including Field
1 were observed in V and I.  The exposure time for each field was
1000s in each filter.  The total area on the sky, excluding a region
around the cluster NGC 1866 and a second smaller cluster, was 622
arcmin$^2$.  Reduction and photometry were done using standard IRAF
tasks.  These data are complete to an apparent magnitude of
$m_V\approx 21$.  Photometric consistency between the ground-based and
WFPC2 data was checked by comparing 37 stars common to both;
differences in V and I were $0.0\pm 0.1$ magnitudes.

\section{COMPARISON BETWEEN THREE LMC FIELDS}
\subsection{Color-Magnitude Diagrams}
Color-magnitude diagrams (CMDs) for the three WFPC2 fields are shown
in Figure 2.  The faintest main sequence turnoff for the three fields
occurs at $M_V\approx 3.5$, and a clear main sequence extends roughly
four magnitudes fainter.  The number of stars in each field is roughly
comparable.  Error bars plotted in Figure 2 are average one $\sigma$
errors as determined by the aperture photometry routine.  Larger
errors in Field 3 are a result of the shorter exposure time.  Major
features in these CMDs, such as the main sequence, the main sequence
turnoff, and the red giant branch, occur at the same magnitude and
color, suggesting similar stellar populations.  We first compare
stellar distributions across the lower main sequence as these are
sensitive to metallicity variations between the three populations.  We
then compare the distributions across the upper main sequence as these
probe variations between the recent star formation histories of the
three fields.

As a statistical test of comparison, we use the one dimensional
Kolmogorov-Smirnov (KS) test.  This test is used to compare the three
fields, as well as to compare the observed luminosity functions to
simulated stellar populations.  The KS test gives the probability (P)
that the deviations between two distributions are the same as would be
observed if they were drawn from the same population.  Two
distributions are considered different if the probability that they
are drawn from the same parent distribution can be ruled out at a
confidence level greater than 95\% ($P\leq 5\%$).  If two
distributions cannot be proved different, we infer that the
populations are similiar, although the KS test does not imply that
these distributions are the same.  We estimate the
sensitivity of this test to minor differences in the star formation
history using simple simulations described below.

As seen in Figure 2, stars appear to be concentrated towards the blue
side of the lower main sequence.  The lower main sequence of Field 1
appears the most concentrated towards the blue, Field 3 appears the
least concentrated.  For Field 3, this is likely the result of a lower
exposure time, but for Field 2 may reflect a real difference with the stellar
population in Field 1.  The histograms of Figure 3 plot the color
distributions for Fields 1 and 2 in several magnitude bins across the
lower main sequence.  The total number of stars in each histogram has
been normalized to the number in Field 1. We used the KS test to
statistically compare the distribution of stars across the lower main
sequence in these fields.  In Figure 3, the relatively high values of
the KS probability, P= 0.2, 0.3 and 0.9, indicate that we cannot
demonstrate that the two samples are drawn from different populations.

The signal-to-noise in Field 3 is lower due to a shorter exposure
time.  In order to compare lower main sequence distributions, we add
gaussian noise to the higher signal-to-noise Field 1 using the average
errors from Field 3 and perform the KS test.  In Figure 4, the
distribution of lower main sequence stars in Field 3 is compared to
the Field 1 plus noise distribution.  Again, we cannot prove that the
distributions are different at a high confidence level.  The
application of the KS test to the lower main sequence color
distributions is weakened by the small number of stars in each
comparison, however, it does suggest similar stellar populations in
the three field.

Although the distribution of stars will also be influenced by age
variations and the presence of binaries, the similarity in the mean
colors of the lower main sequences suggest that the three fields have
similar mean metallicities.  To estimate the sensitivity of our tests
to differences in metallicity, we use stellar models described
below to simulate stellar populations with identical star formation
histories but different metallicity distributions.  We find that the
KS test is sensitive to metallicity differences if, between the two
populations, at least 25\% of the stars have a factor of four
difference in metallicity.  This fraction decreases to 20\% of stars
if the metallicity differs by a factor of ten.  These results are
robust for several different assumed star formation histories.

From the width of the lower main sequence, we can rule out the
possibility that the observed stellar populations have a single
metallicity.  The standard deviations of the observed lower main
sequence distributions shown in Figure 3 are $\sigma_{(V-I),observed}=
0.06, 0.07, 0.08$ for the three magnitude bins.  Field 3 is not
included due to the higher noise.  We compare this width with that of
a simulated single metallicity population, assuming a constant star
formation rate from 12 Gyr to the present and a 50\% binary fraction.
Observational errors are simulated by randomly sampling the error
distributions determined from artificial star tests, and include both
systematic and random errors.  Independent of assumed metallicity, the
standard deviation of the lower main sequence was $\sigma_{(V-I),
model}=0.05, 0.05, 0.06$ for the same magnitude bins.  We find that
the difference between the observed and model variances is significant
based on a F-test.  Populations having a smaller age range or a lower
assumed binary fraction will have narrower distributions.  Therefore,
the observed lower main sequence is wider than is expected for a
single metallicity population.  Since this is a differential
comparison, we expect it to hold regardless of the adopted stellar
models.  The derivation of the absolute metallicity is discussed in
$\S 4.1$ and may be model dependent.

The distribution of stars across the upper main sequence is shown in
Figure 5.  The width of the main sequence is broader than would be
expected from a single age population.  From stellar evolution models,
stars brighter than $M_V\approx 3$ evolve steadily redward during
their main sequence lifetimes.  Thus, a uniform distribution across
the upper main sequence suggests no intense star formation bursts have
occurred during the lifetimes of these more massive stars.  Gallagher
\etal \markcite{gall96}(1996) show that the distribution of upper main
sequence stars in Field 1 is roughly consistent with a star formation
rate constant for the past 3 Gyr.  The KS test gives no evidence that
the distributions of stars across the upper main sequence in the three
fields are different.  This test is not sensitive to variations in the
star formation history more recent than 0.1 Gyr, due to small number
statistics at the brightest magnitudes.

A second distinct main sequence turnoff discussed by Gallagher \etal
\markcite{gall96}(1996) and seen near $M_V\approx 2.5$ in Figure 2a is
not observed in the remaining two fields.  Gallagher \etal interpret
this turnoff and associated excess of stars at $M_V=2$ in the
Hertzsprung gap as signatures of a short star formation burst
occurring $\sim 2$ Gyr ago.  Although many stars populate this region
of the CMDs in Field 2 and 3, the lack of a single distinct subgiant
branch in these fields argues against a short ($t\leq 0.1$ Gyr),
global 2 Gyr burst throughout the LMC.  The subgiant excess observed
in Field 1 may arise from a statistical fluke, the remnants of a
localized star formation burst, or a dissolved cluster.

\subsection{Comparison of Luminosity Functions}
The observed, uncorrected, differential luminosity functions are
shown
in Figure 6.  We perform the KS test between Fields 1 and 2
over the
range $-0.5\leq M_V\leq 7.5$ and between Field 1 and Field 3 over
the
range $-0.5\leq M_V\leq 6$.  The resulting KS probabilities
between
fields are $P_{LF_{Field 1-2}}= 0.37$ and $P_{LF_{Field
1-3}}=0.26$,
thus we are not able to show that the three luminosity functions
are
different.

\subsection{Summary of Comparison}
We conclude, from analysis of the luminosity function and distribution
of stars across the main sequence, that the three observed regions in
the LMC field contain statistically indistinguishable stellar
populations.  In addition, we have calculated R-ratios as defined by
Bertelli \etal \markcite{bmcb92}(1992) and discussed below ($\S 4.2$),
which compare the number of stars in different evolutionary phases;
these are shown in Table 2.  These ratios are also the same between
the three fields, within the errors determined by number statistics.

We cannot prove that the star formation histories in the three fields
are different; however, this does not imply that they are
identical. To estimate the sensitivity of our tests to variations in
star formation history, we simulate a simple star formation history
with a constant star formation rate.  We compare this to models in
which star formation is turned off completely for short lengths of
time at different epochs. We find that our statistical tests are
unable to conclusively distinguish between such models for variations
in star formation rate over periods shorter than 1 Gyr anytime in the
past 4 Gyr, or over periods shorter than 2 Gyr anytime before 4 Gyr
ago.

We next
compare our observations to stellar
models using similar statistical tests in order to place
constraints on possible
star formation histories.  As we have shown that our
methods cannot distinguish between the three stellar populations,
we combine the observations of the three fields to improve number
statistics.

\section{STAR FORMATION HISTORY}
\subsection{Stellar Models}
To determine the star formation history in the three fields, we
compare our observations to simulations made using the stellar
evolution models published by the Padua group (Bertelli \etal
\markcite{bbcf94}1994 and references therein).  These isochrones range
in metallicity from $0.0004\leq Z\leq0.05$ (-1.7$\leq[Fe/H]\leq$0.4)
and are calculated for stellar masses down to $0.6M_{\odot}$.
Depending on the metallicity, this corresponds to an absolute
magnitude of $M_V\simlt 8$, roughly the magnitude limit of our
observations. The Padua models are calculated with mild convective
overshoot and the most recent Livermore group radiative opacities
(Iglesias \etal \markcite{irw92}1992).  UBVRI magnitudes for these
models have been calculated by Bertelli \etal \markcite{bbcf94}(1994).
There is evidence for nonsolar abundance ratios in the LMC, such that
the $\alpha$-element are enhanced relative to the solar ratio (Luck \&
Lambert \markcite{ll92}1992).  Although $\alpha$-enhanced models are
not currently available, Salaris \etal \markcite{scs93}(1993) have
found that under some conditions, $\alpha$-enhanced isochrones are
well mimicked by scaled solar metallicity isochrones.  The effect of
$\alpha$-element enhancement is to shift a solar abundance isochrone
towards the red, which would lead to an overestimate of our derived
metallicities.

Comparing the observed CMDs to single age isochrones, we find the blue
edge of the upper main sequence is best matched by an isochrone of
metallicity Z=0.008, whereas the red giant branch can be fit by either
an old to intermediate age, metal poor isochrone ($t\geq 2$ Gyr and
Z=0.0004) or a young, higher metallicity isochrone ($t\leq 2$ Gyr and
Z=0.001).  This point is well illustrated in figure 4 of Holtzman
\etal (1997).  Problems with the stellar atmospheres at lower
temperatures and/or the evolutionary models of giant stars may be
responsible for the apparent mismatch. Alternatively, it may be
related to our use of solar abundance ratio isochrones. Because of
these possible problems, we do not use the color of the giant branch
to derive stellar population parameters.  However, some of the derived
parameters, in particular metallicities, are sensitive to the model
colors of main sequence stars.  We note that our constraints on such
parameters are derived assuming that these stellar models are perfect;
we allow for random errors in the observations but not for systematic
errors in the models.

\subsection{Simulations}
The presence of a bright main sequence, seen in Figure 2, suggests
recent star formation activity, whereas the faintest main sequence
turnoff at $M_V\approx 3.5$ implies an older star formation epoch.  We
therefore simulate CMDs for a mixture of simple stellar populations.
Throughout the simulations, we assume that 50\% of the stars are
binaries with uncorrelated masses (Reid \markcite{reid}1991). In order
to preserve isochrone shape during age interpolation, each isochrone
is resampled into 100 equally spaced mass points within each of 9
evolutionary epochs.  No interpolation is made in metallicity.  The mass and age of each star are chosen
according to an input initial mass function and star formation
history.  The absolute magnitude and color of each star can be
determined for any desired age by interpolating point by point over
the resampled isochrones.  An apparent magnitude is determined based on the extinction and distance modulus.  The star is
considered detected or rejected according to the completeness
histograms calculated during the artificial star tests described
above.  An observational error is given to each detected star by
randomly sampling the error distribution appropriate for the star's
magnitude determined from the artificial star tests (see $\S 2.1$).
These errors include both random and systematic effects.

The number of free parameters in determining a star formation history
is large.  Our observations do not provide enough constraints to
justify an exhaustive search of parameter space.  Instead, we choose
to discuss six representative formation histories: constant star
formation throughout the history of the LMC, the two preferred
histories of Bertelli \etal \markcite{bmcb92}(1992) and Vallenari
\etal \markcite{vcba96b}1996b, two proposed histories of Holtzman
\etal \markcite{holt97}(1997), and a formation history motivated by
the observed age distribution of LMC globular clusters.  In all cases,
we assume the age of the oldest LMC stars to be 12 Gyr based on age
estimates of the oldest LMC globular clusters (van den Bergh
\markcite{vand91}1991).  The parameters used in each simulation are
shown graphically in Figure 7.

To compare observed and theoretical stellar populations, we use a
combination of two methods.  First, we compare observed and model
luminosity functions using the 1-D KS test as described in $\S 3.1$.
The three fields are combined to create the observed luminosity
function and are compared with models over a range $-0.5\leq M_V\leq
6$.  A star formation history is considered acceptable if the
probability, P, that the luminosity function is drawn from the same
population is greater than 5\%.  The sensitivity of this test to
variation in the star formation history is the same as that discussed
in $\S 3.3$.

In comparing luminosity functions without regard to color, some star
formation history information is lost, especially for stars brighter
than the main sequence turnoff.  Thus, in addition to luminosity
function fitting, we use the R-method described in detail by Bertelli
\etal \markcite{bmcb92}(1992) for $M_V\leq 3$.  Briefly, this method
defines three stellar number ratios, each sensitive to different
parameters in the star formation history.  The first ratio is defined
as:

\begin{equation}  R_1 =
\frac{\#~of~main~sequence~stars}{\#~of~red~giant~stars},~~M_V
\leq 3
\end{equation}

\noindent The separation between main sequence and red giant stars is
determined by the lines shown in Figure 2, and is consistent
throughout the analysis.  The next two ratios compare the number of
bright to faint stars on the main sequence and the red giant branch.
The magnitude separating the upper and lower regions is defined at
$M_V=1.5$, chosen to be below the red clump.  The ratios are defined
as:

\begin{equation}  R_2 =
\frac{\#~of~upper~red~giant~stars~(M_V\leq
1.5)}{\#~of~lower~red~giant~stars~(1.5\leq M_V\leq3)}
\end{equation}

\begin{equation}  R_3 =
\frac{\#~of~lower~main~sequence~stars~(1.5\leq
M_V\leq3)}{\#~of~upper~main~sequence~stars~(M_V \leq 1.5)}
\end{equation}

The R-ratios depend, in different ways, on the slope of the initial
mass function and the relative number of young and old stars; see
Bertelli \etal (1992) for a more extended discussion of these
dependencies.  The R-ratios for the three observed fields, as well as
the ratios resulting from combining the three fields, are shown in
Table 2.  The errors in this table are one $\sigma$ errors as
determined from number statistics.

The WFPC2 fields do not provide representative numbers of stars at the
brightest magnitudes due to saturation and the small field of view.
We corrected for this by combining the \hst data with ground-based
data covering a significantly larger field and including Field 1.  A
more detailed analysis of these data was presented by Stappers \etal
\markcite{stap97}(1997).  To combine star counts from ground and
space-based data directly, we applied a scale factor determined by the
ratio of observed areas.  We use ground-based star counts for
magnitudes brighter than $M_V=1.5$ to recalculate the R-ratios of the
combined field.  As compared to the uncorrected ratios (Table 2,
second to last row), the use of ground-based data results in an
average 10\% corrections to the R-ratios (Table 2, last row).
Constraints from the R-ratios, particularly $R_2$, may be
less secure than those based on the main sequence because of the
larger uncertainties in modelling later phases of stellar evolution.

\subsection{Star Formation History}
We compare the observed luminosity function and R-ratios with those
calculated for the six star formation histories shown in Figure 7.
The observed and computed R-ratios, as well as the KS probability
resulting from a comparison of the observed and model luminosity
functions, are given in Table 3.  The observed and simulated luminosity
functions for each model are shown in Figure 8.  The simulated
luminosity functions are normalized to match the observations at
$M_V=4$.

The simplest star formation history tested assumes a Salpeter IMF and
a constant star formation rate since the formation of the LMC 12 Gyr
ago (Figure 7a).  In examining the parameter $R_1$, this simple
formation scenario does not produce enough bright main sequence stars
by more than a factor of two relative to the number of observed evolved stars.
This deficiency has motivated most modelers to include an enhancement
in the recent star formation rate.

Bertelli \etal \markcite{bmcb92}(1992) observed a field $17'$
southwest of Field 1.  Using the R-method, they derive a star
formation history in which the star formation rate was initially low
and then increased by a factor of ten 4 Gyr ago, as shown in Figure
7b.  Although this model reproduces the observed R-ratios reasonably
well, it does not match the observed luminosity function.  As shown in
Figure 8b, this formation scenario produces too many bright ($M_V \leq
3$) stars relative to faint stars.  We note that the 1992 Padua
stellar models used to derive this history allow for more convective
overshoot than the 1994 models used in this paper.  The use of more
recent models decreases the inferred time when the star
formation increase rate began to approximately 2 Gyr ago (Bertelli, private
communication).  Such a star formation history was used by Vallenari
\etal (1996b) to match observations in several of their fields and its
luminosity function is shown in Figure 8c.  It produces even more
bright stars relative to the number of observed faint stars, and is
ruled out by our observed luminosity function.  

In their simulations,
Vallenari \etal allow for interpolation in metallicity, whereas our
models use discrete metallicity isochrones.  We find that the
luminosity functions of the three individual metallicities which
contribute to the final formation scenario are inconsistent with the
observations in the same direction as the composite Vallenari \etal
luminosity function.  Therefore, this simplification is not the source
of discrepancy between the Vallenari \etal scenario and our
observations.

Holtzman \etal \markcite{holt97}(1997) find that a steeper IMF slope
is necessary in order for a 4 Gyr, ten-fold star formation rate
increase to match the observed luminosity function.  For an IMF slope
$\alpha=-2.75$, this star formation history (Figure 7d) is consistent
with the observed luminosity function, but not with the R-ratios
observed in our fields.

Of published star formation histories for the LMC field, the only
one which reproduces both the luminosity function and
R-ratios of our observation is that of Holtzman \etal
\markcite{holt97}(1997).  In this formation scenario, the star
formation rate remains constant for
a majority of the LMC history and is increased by a factor of
three from 2 Gyr ago to the present; the
star formation rate is also slightly higher for the oldest stars
(Figure 7e).  In contrast to the
previous three
scenarios which produce primarily young stellar populations, this
star formation history produces a population with roughly equal
number of stars older and younger than 4 Gyr.  The simulated CMD
for this star formation history is shown in the right panel of
Figure 9.

A star formation history based on the age distribution of LMC globular
clusters cannot fit the observed field population.  The age
distribution of LMC clusters is bimodal (Figure 7e) with $\geq 90\%$
of clusters formed between $10^6$ years and 3 Gyr ago, and $\leq 10\%$
of clusters having ages between 10 and 12 Gyr (van den Bergh
\markcite{vand91}1991).  There are almost no known intermediate age
(3-10 Gyr) clusters in the LMC (Girardi \etal \markcite{gira95}1995;
however see Sarajedini \etal \markcite{sara95}1995), however, an
intermediate population is necessary to reproduce our observations.
If no clusters have been destroyed, we conclude that the star
formation history of LMC globular clusters is not mimicked by the
field population.

\subsection{Chemical Evolution}
It is possible to predict the chemical history of the LMC from its
star formation history using the simple closed box model of chemical
evolution.  We assume a one-zone evolution model with no infall or
outflow, zero initial metal content and instantaneous recycling
(Searle \& Sargent \markcite{ss72}1972).  This model has successfully
predicted the relationship between metallicity and current gas
fraction in Magellanic irregular galaxies, although it has less
success predicting this relationship in larger spiral systems (Binney
\& Tremaine \markcite{bt87}1987).  We assume a present day gas to
total LMC mass ratio of $M_{gas}/M_{total}=0.2$ (Cohen \etal
\markcite{cohe88}1988) and an effective yield of $p=0.005$, chosen so
that the present day metallicity matches that inferred from the upper
main sequence ($Z=0.008$).  We compare the predicted chemical
evolution for two star formation histories in Figure 10.

In the upper panel of Figure 10, the Holtzman \etal \markcite{holt97}(1997) star
formation history suggests that the metallicity in the LMC has doubled
in the past 2 Gyr.  The Vallenari \etal (1996) formation history
(bottom panel) implies a factor of five metallicity increase in the
past 2 Gyr.  The chemical evolution predicted by the Holtzman \etal
model is consistent with planetary nebula observations by Dopita \etal
\markcite{dopi97}(1997) which suggest the metallicity of the LMC has
almost doubled in the last 2 Gyr.

We also note that a significant metal poor population is predicted by
the closed box model, regardless of the details of the star formation
history.  For an effective yield of $p=0.005$, the closed box model
predicts 22\% of stars have metallicities less than Z=0.001.  This
fraction is inversely proportional to the assumed yield.

Simulated lower main sequences suggest a similar fraction of LMC field
stars are metal poor.  Lower main sequence cross sections are shown
for two star formation histories in Figure 11.  In the left column,
Vallenari \etal's model assumes a metallicity range Z=0.008-0.001.
These distributions are displaced to the red of the observations by as
much as a tenth of a magnitude and are significantly narrower.
Although some redward evolution occurs for low mass stars during their
main sequence lifetime, changing the star formation history alone is
not enough to explain this color shift. The Holtzman \etal model is
better able to fit the lower main sequence, as it includes an old,
metal poor component as shown in the right column of Figure 11.  In
this simulated population, 20\% of stars have a metallicity Z=0.0004
([Fe/H]= -1.7).  None of the models considered here, however, include
chemical evolution in a fully self-consistent manner.

Direct measurements of metallicity in the LMC field have been limited
to bright stars, but possibly suggest a similar fraction of metal poor
stars.  Olszewski \markcite{olsz93}(1993) spectroscopically determined
metallicities for 36 red giant stars in an outer LMC field near NGC
2257 and found that 8 or 9 ($\sim 25\%$) of these stars had
metallicities below Z=0.001 ([Fe/H]= -1.3).  More metallicity
observations are needed to determine the size of a metal poor
component in the LMC field (Suntzeff \markcite{sunt97}1997).

\section{SUMMARY}
We present deep WFPC2 observations of three fields in the outer disk
LMC.  We find no conclusive evidence for variation in the stellar
populations between the three fields based on the morphologies of the
color-magnitude diagrams, the luminosity functions, and the relative
numbers of stars in different evolutionary stages.  

In apparent contrast to our results, Vallenari \etal
\markcite{vcba96b}(1996b) find significant variations in the star
formation history correlated with azimuthal angle in the LMC field.  A
direct comparison with their results, however, is difficult.  The
R-ratios of our Field 1 agree with those calculated for the nearly
overlapping Vallenari \etal NGC 1866 field.  Field 3 is located reasonably
close to Field 1, therefore the similarity of this field to Field 1 provides
no direct contradiction with the Vallenari \etal results. A direct discrepancy
comes from Field 2, which is located $\sim 1^{\circ}$ from the
Vallenari \etal field LMC-61.  We find significant disagreement in the
ratio $R_3$ between these two fields.  This discrepancy may be due to
the small difference in location or to some systematic error in one or
both of the samples.  A much larger survey is required to determine whether a correlation exits between the star formation history and
position in the LMC field (Zaritsky, Harris \& Thompson
\markcite{zht97}1997).

Other evidence that the star formation history varies within the LMC
comes from Elson \etal \markcite{egs97}(1997), who present evidence
that the stellar population in the LMC bar is different from those
presented in this paper.  They analyze \hst observations for a field
in the bar of the LMC and identify additional peaks in the color
distribution between $20.0 \leq m_V \leq 22.5$ not associated with the
red giant branch or the most recent epoch of star formation.  They
attribute these peaks to a burst of star formation between 1 and 2
Gyr depending on the assumed metallicity.  We do not find evidence for
this population in our observed color distribution.  Elson \etal
associate this population with the formation of the LMC bar.

We have compared our observations with stellar models to place constraints
on possible star formation histories in the three fields. These constraints
are an improvement over previous results as they incorporate both \hst
and ground-based data, allowing measurements of the deep main sequence
luminosity function, the distribution of stars in the upper main
sequence band, and the relative number of bright stars which probe
different evolutionary phases.  Of previously considered star
formation histories, the only one which is consistent with all of our
observations has a star formation rate which is roughly constant for
10 Gyr, then increases by a factor of three for the past 2
Gyr. Contrary to many previous models, this produces a population
which is {\it not} dominated by young stars.  Although the star
formation history of the LMC is clearly more complicated, this simple
picture should provide a useful guide to understanding the formation
of our nearest neighbor.

This work was supported in part by NASA under contract NAS7-918 to
JPL and a grant to M.G. from the New Mexico Space
Grant Consortium.

\newpage

\figcaption{Image of the LMC showing the
approximate positions of the three \hst fields.  Field
1 is off the image roughly the length of the arrow in the
direction indicated.   North is up, East is to
the left.  Taken from Sandage \protect \markcite{sand61}(1961).
\label{fig1}}

\figcaption{Color magnitude diagrams for
Field 1, 2 and
3.  One $\sigma$ error bars are shown as determined by the
aperture photometry routine.  The vertical solid line is the
boundary between main
sequence
and evolved stars used to determine the R-ratios.  Horizontal
lines at $M_V=3$ and 1.5 denote the faint magnitude limit
and the separation between bright and faint stars used in the
R-method, respectively.
\label{fig2}}

\figcaption{Histogram of stars across the lower
main sequence for
three magnitude bins.  The solid histogram corresponds to Field
1, the dotted to Field 2.  The distributions are statistically
indistinguishable as shown by
relatively high values of P, the
KS probability.  The number of
stars in
Field 2 have been normalized to Field 1.  The error bars show
one $\sigma$ errors.
\label{fig3}}

\figcaption{Comparison of lower main sequence
distributions
between Field 1 (solid histogram) and Field 3 (dash-dotted
histogram). 
For comparison, gaussian noise has been added to Field 1.
\label{fig4}}

\figcaption{Distribution of stars across the upper
main
sequence. 
In the first column, Field 1 (solid histogram) is compared with
Field
2 (dotted),
in the second column, Field 1 (solid) is compared to Field 3
(dash-dotted). At these magnitudes,
photometry errors in the three fields are much smaller than the
bin size. 
\label{fig5}}

\figcaption{The three observed luminosity
functions.  Field 3 has been normalized arbitrarily. 
\label{fig6}}

\figcaption{Schematic representation of the six star
formation
histories
tested.  The initial mass slope, $\alpha$, is given in the upper
right corner, the metallicity is shown above each epoch.
\label{fig7}}

\figcaption{Luminosity functions for the six star
formation
histories compared to the observed luminosity function of the
three fields.  Model
luminosity functions are normalized to match the observations at
$M_V=4$.
\label{fig8}}

\figcaption{Left panel: Observed CMD for the combined fields.
Right panel: Simulated CMD resulting from our preferred star formation
history (model (e), Holtzman \etal).
\label{fig9}}

\figcaption{Chemical evolution of the LMC
predicted
by the
closed
box model for two star formation histories.  The model assumes a
present day gas mass to total mass of 0.2 and an average effective yield of
$p=0.005$.
\label{fig10}}

\figcaption{Comparison of the observed lower main
sequence
distribution (solid histograms) to two model distributions.  A
metal poor component is needed to match the observed
color distribution.  
\label{fig11}}

\newpage

\makeatletter
\def\jnl@aj{AJ}
\ifx\revtex@jnl\jnl@aj\let\tablebreak=\nl\fi
\makeatother
\vskip 0.5 in

\begin{center}
\begin{tabular}{c c c c c}
\multicolumn{4}{c}{\bf Table 1.} \\
\multicolumn{4}{c}{ Summary of Observations} \\ \hline \hline
\multicolumn{1}{c}{Field} &
\multicolumn{1}{c}{$\alpha_{2000}$} &
\multicolumn{1}{c}{$\delta_{2000}$} &
\multicolumn{1}{c}{Exposure time} & \\ \hline
Field 1  & $5^h 14^m 44^s$ & $-65^{\circ} 17' 43''$ & 4000s  \\ 
Field 2  & 5 58 21 & -68 21 18 & 2500s  \\ 
Field 3  & 5 4 14 & -66 35 2.4 & 1000s \\ \hline

\end{tabular}
\end{center}

\begin{center}
\begin{tabular}{c c c c c}
\multicolumn{4}{c}{\bf Table 2.} \\
\multicolumn{4}{c}{ Observed R-Ratios} \\ \hline \hline
\multicolumn{1}{c}{Field} &
\multicolumn{1}{c}{$R_1$} &
\multicolumn{1}{c}{$R_2$} &
\multicolumn{1}{c}{$R_3$} & \\ \hline 

Field 1  & $ 5.63\pm 0.89 $& $1.80\pm 0.70$ & $ 3.49\pm 0.68$\\ 
Field 2  & $ 4.30\pm 0.73 $& $1.72\pm 0.63$ & $ 3.20\pm 0.64$\\ 
Field 3  & $ 4.81\pm 0.72 $& $1.89\pm 0.63$ & $ 3.63\pm 0.63$\\ 
Combined Fields  & $4.84\pm 0.34$ & $1.81\pm 0.24$ & $3.46\pm
0.24$ \\ 
\hline
Combined + Ground & $4.24\pm 0.18$ & $2.22\pm 0.24$ & $3.43\pm
0.12$ \\ \hline
\end{tabular}
\end{center}

\begin{center}
\begin{tabular}{c c c c c c}
\multicolumn{5}{c}{\bf Table 3.} \\
\multicolumn{5}{c}{ Model star formation results} \\ \hline
\hline
\multicolumn{1}{c}{Model} &
\multicolumn{1}{c}{$R_1$} &
\multicolumn{1}{c}{$R_2$} &
\multicolumn{1}{c}{$R_3$} &  
\multicolumn{1}{c}{$P_{LF}$} & \\ \hline 
Observed Values & $4.24\pm 0.18$ & $2.22\pm 0.24$ & $3.46\pm
0.12$ &\\
 \hline
Constant Star  &  $1.97\pm 0.05$ & $ 1.02\pm0.04$ & $4.02\pm0.14$
&  $9\times10^{-9}$\\ 
  Formation       &  &  & &\\
Bertelli \etal &  $4.00\pm 0.08$ & $ 1.90\pm0.07$ & $3.84\pm0.09$
&  $2\times10^{-11}$\\ 
  (1992)          &  &  & &\\ 
Vallenari \etal &  $5.55\pm 0.12$ & $3.39\pm0.15$ & $2.98\pm0.06$
& $\leq10^{-11}$ \\
  (1996b)          &  &  & &\\
Holtzman \etal &  $3.06\pm 0.08$ & $1.45\pm0.07$ & $4.49\pm0.15$
& 0.14 \\
(1997) $\alpha=-2.75$          &  &  & &\\
Holtzman \etal &  $4.02\pm 0.09$ & $2.08\pm0.09$ & $3.41\pm0.08$
& 0.17 \\
(1997) $\alpha=-2.35$          &  &  & &\\
LMC Cluster    &  $9.53\pm 0.17$ & $3.32\pm0.12$ & $2.21\pm0.03$
&
$\leq10^{-11}$ \\
Age Distribution        &  & & &\\ \hline

\end{tabular}
\end{center}

\begin{figure}[b!] % fig 2
\centerline{\hbox{\epsfig{file=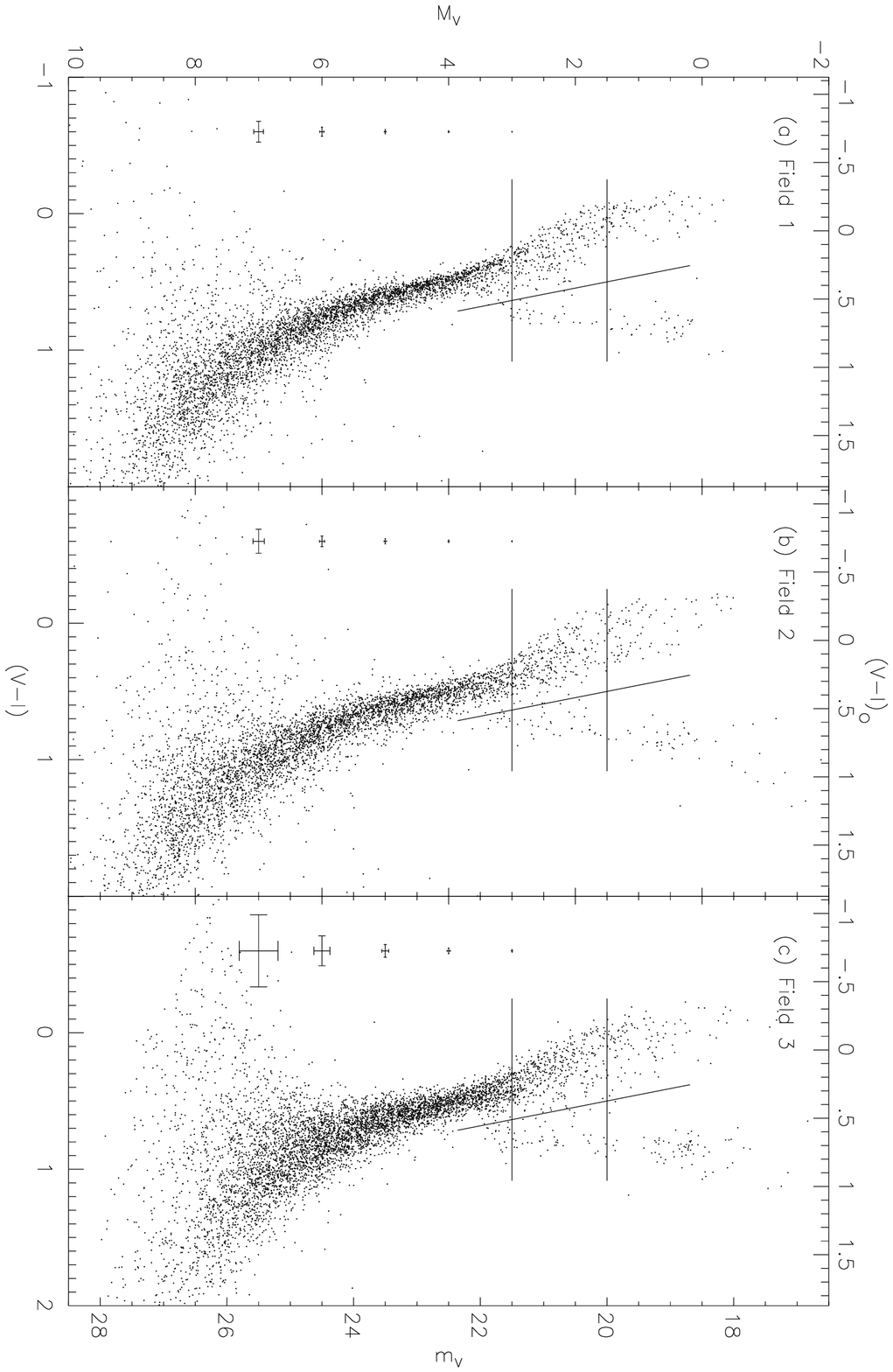, height=6.4 in, width=5. in }}}
\vskip 0.5 in
{Fig. 2.-}
\end{figure}
\begin{figure}[b!] % fig 3
\centerline{\hbox{\epsfig{file=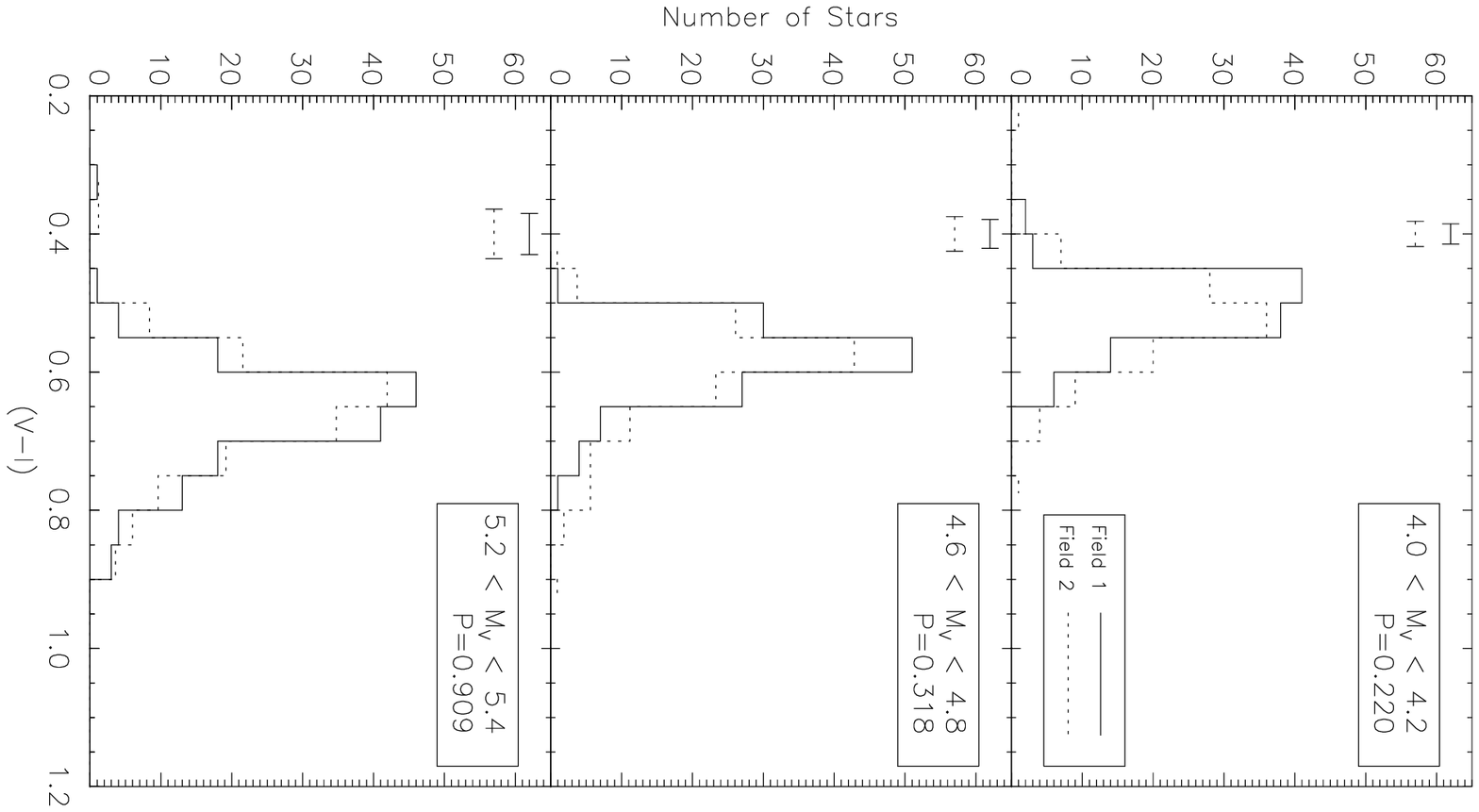, height=6 in, width=6 in }}}
\vskip 0.5 in
{Fig. 3.-}
\end{figure}
\begin{figure}[b!] % fig 4
\centerline{\hbox{\epsfig{file=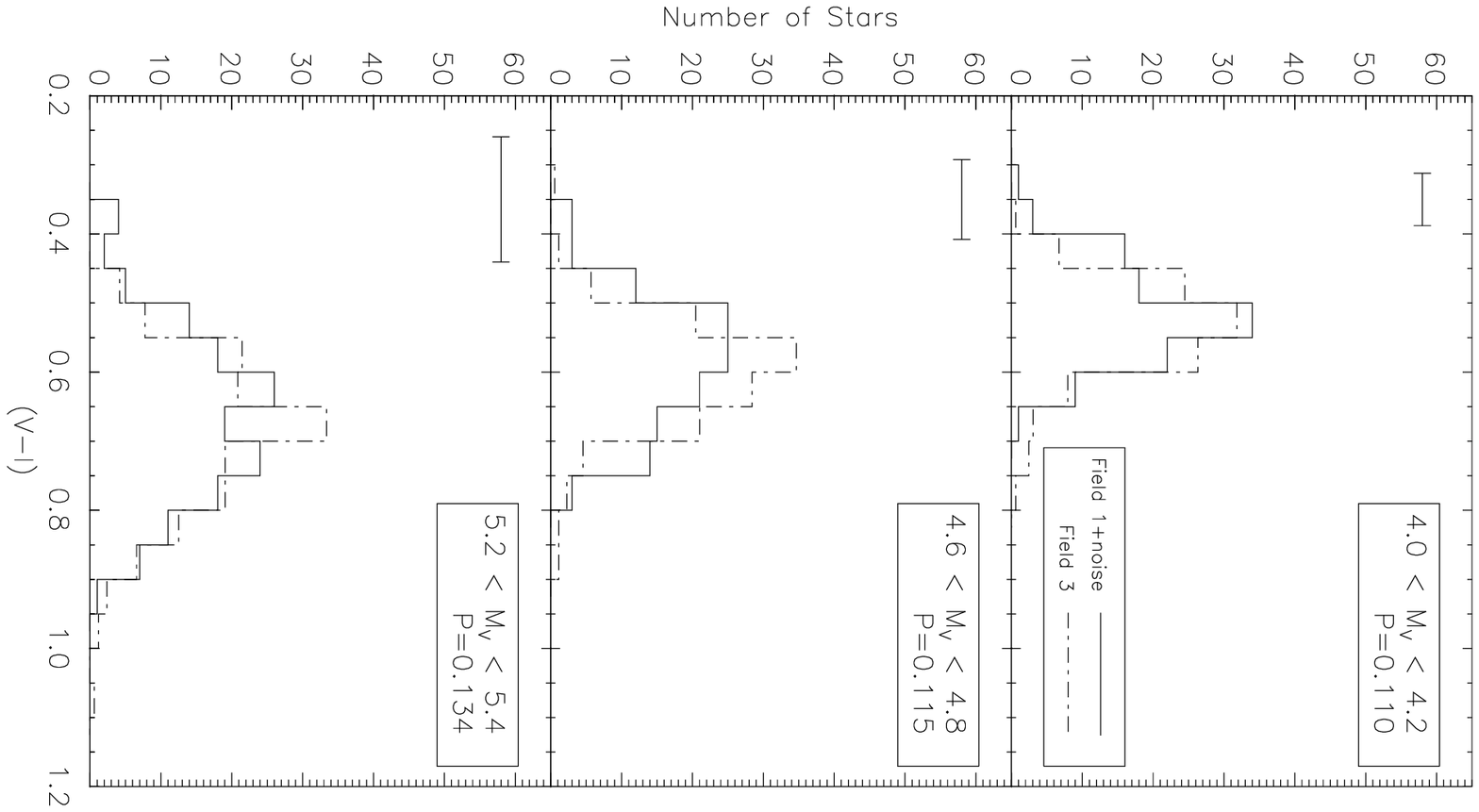, height=6 in, width=6 in }}}
\vskip 0.5 in
{Fig. 4.-}
\end{figure}
\begin{figure}[b!] % fig 5
\centerline{\hbox{\epsfig{file=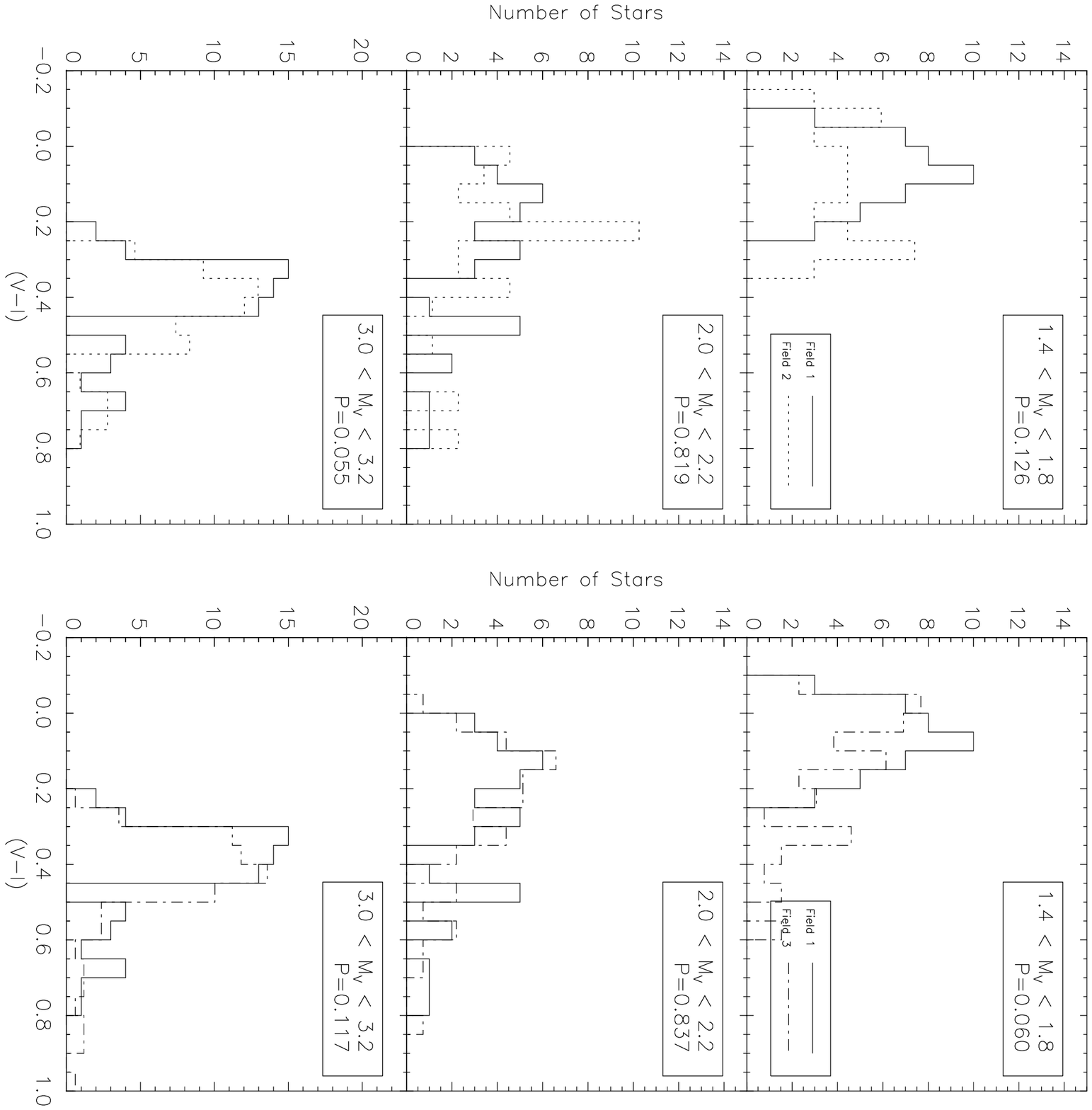, height=6 in, width=6 in }}}
\vskip 0.5 in
{Fig. 5.-}
\end{figure}
\begin{figure}[b!] % fig 6
\centerline{\hbox{\epsfig{file=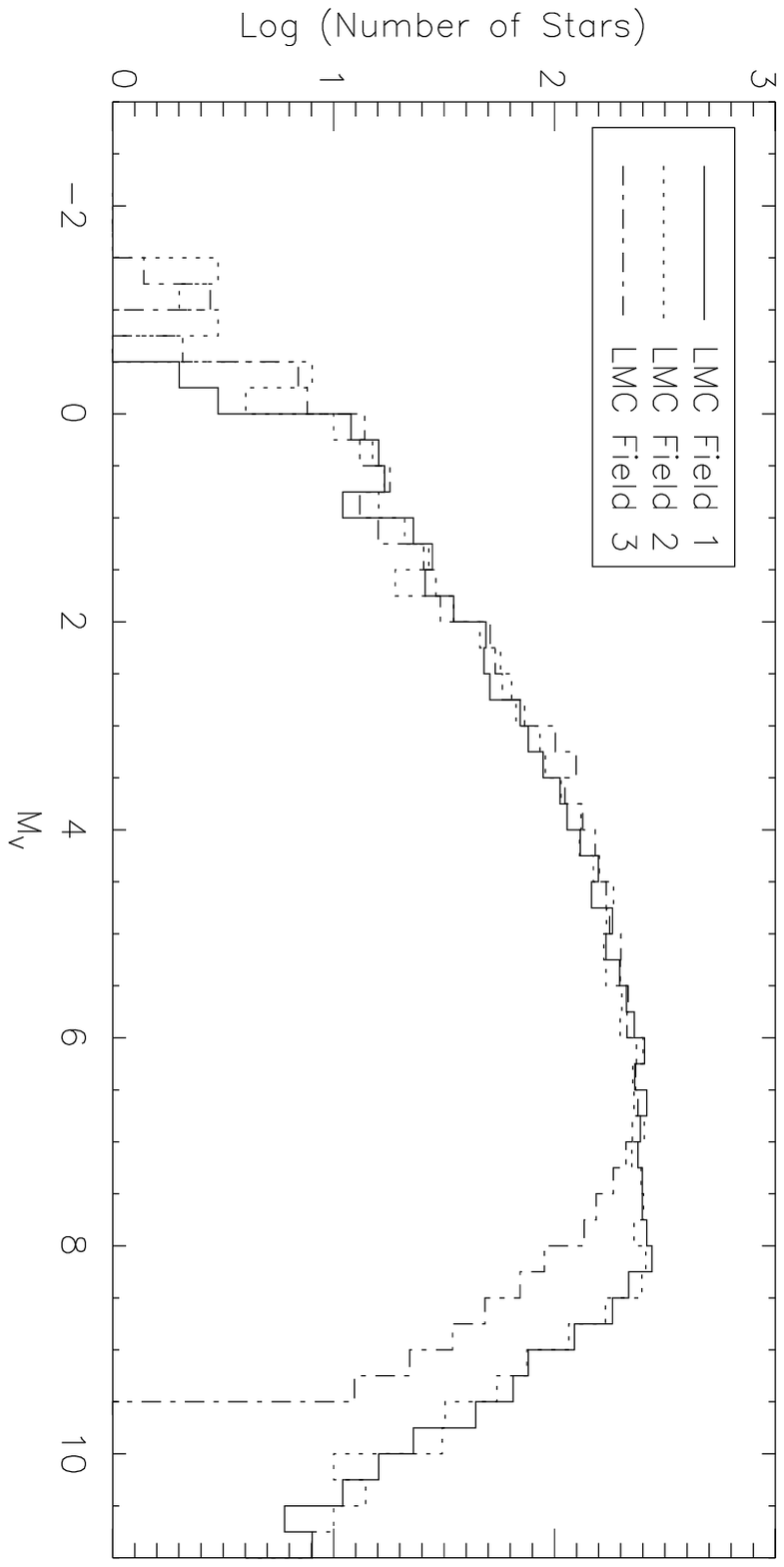, height=6 in, width=6 in }}}
\vskip 0.5 in
{Fig. 6.-}
\end{figure}
\begin{figure}[b!] % fig 7
\centerline{\hbox{\epsfig{file=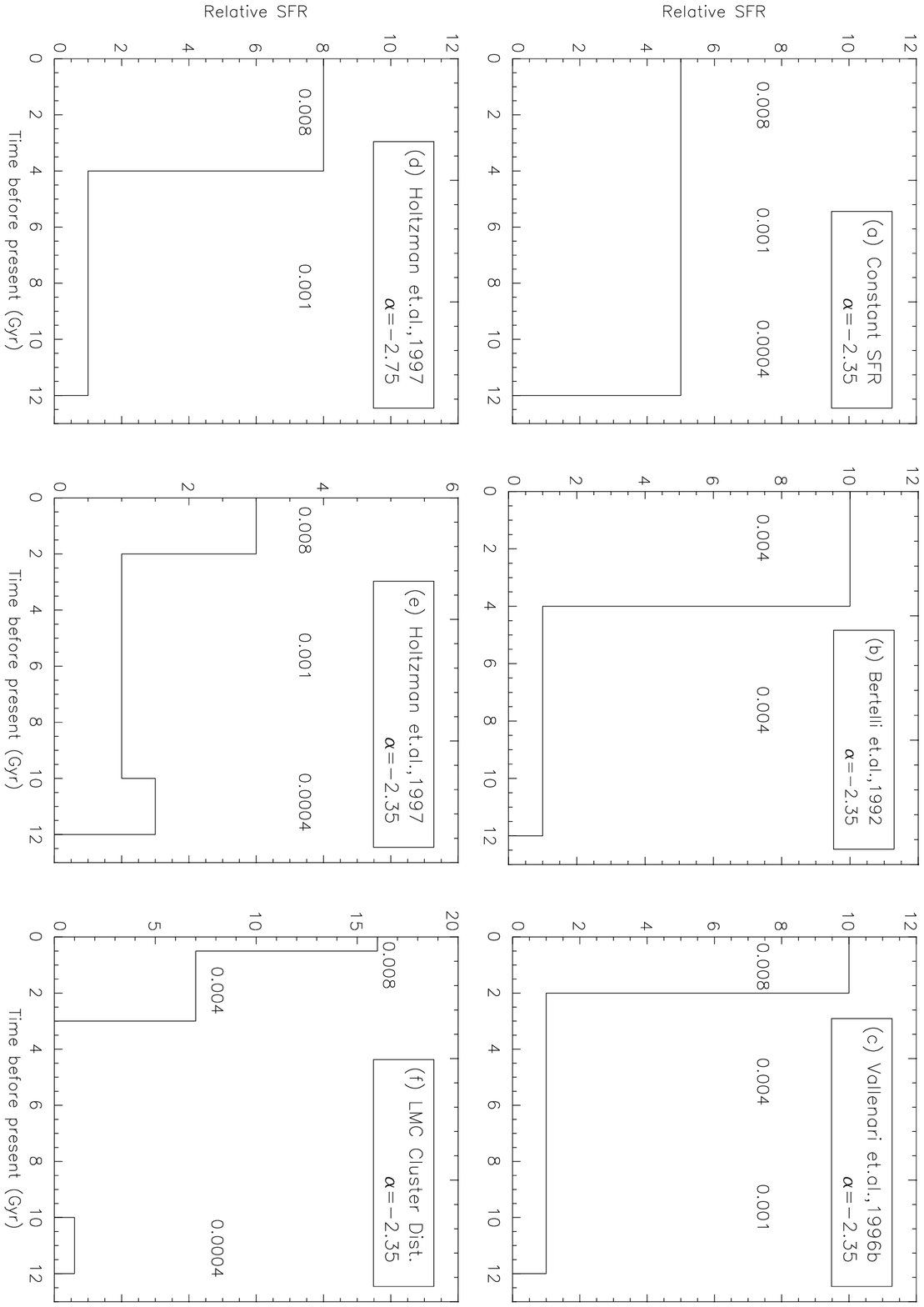, height=6.3 in, width=6 in }}}
\vskip 0.5 in
{Fig. 7.-}
\end{figure}
\begin{figure}[b!] % fig 8
\centerline{\hbox{\epsfig{file=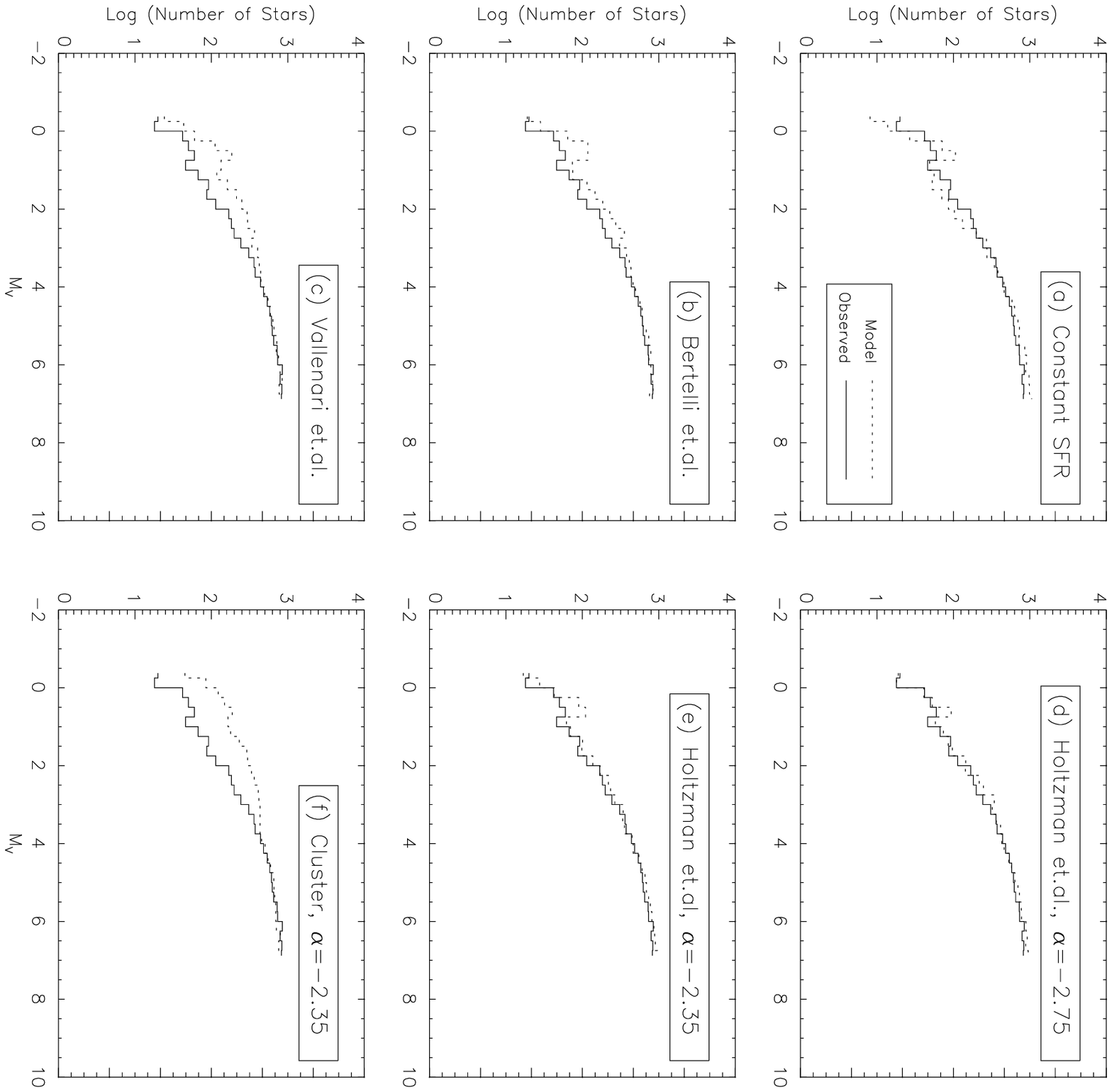, height=6.3 in, width=6 in }}}
\vskip 0.5 in
{Fig. 8.-}
\end{figure}
\begin{figure}[b!] % fig 9
\centerline{\hbox{\epsfig{file=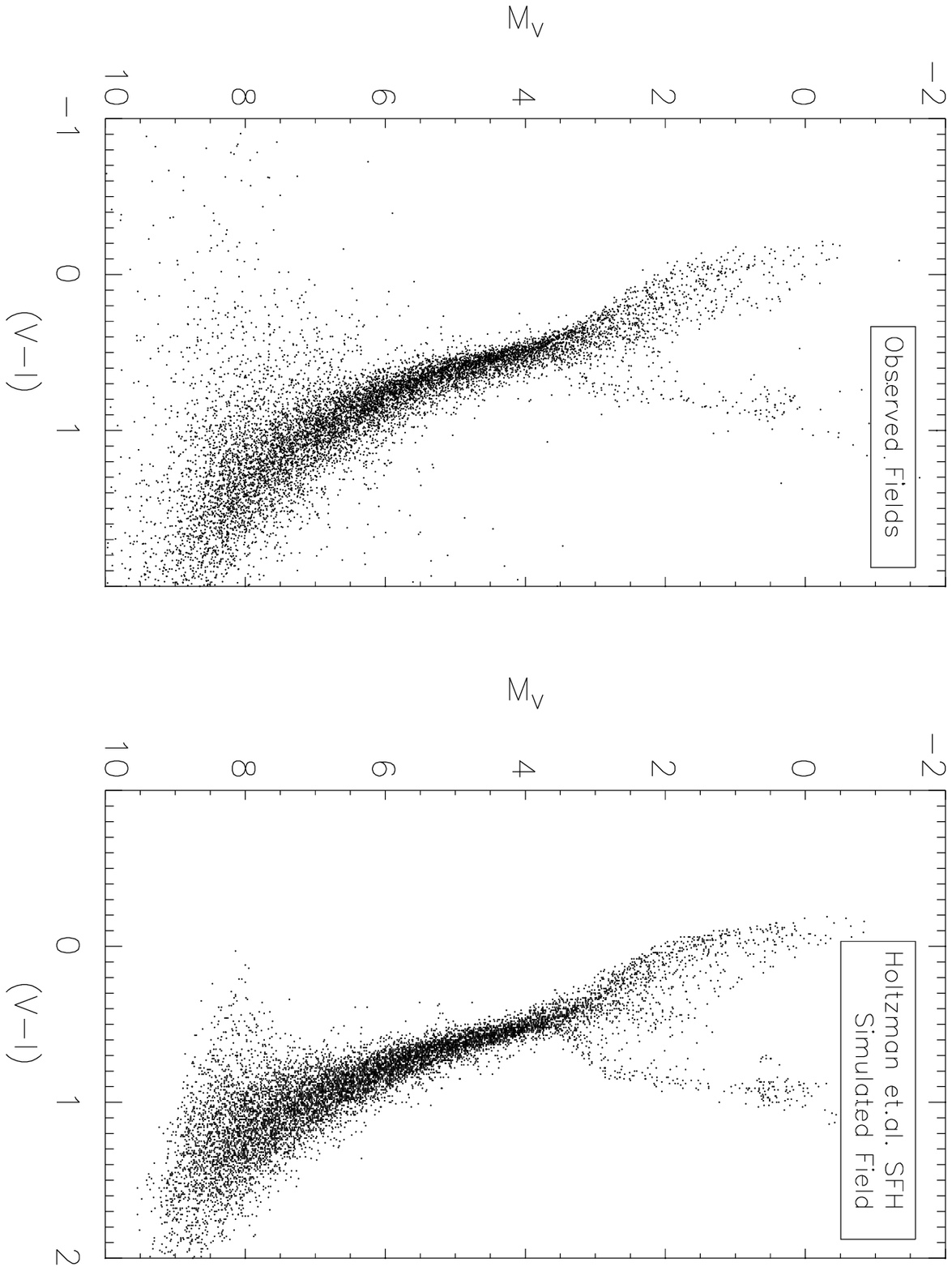, height=6.4 in, width=5.3 in }}}
\vskip 0.5 in
{Fig. 9.-}
\end{figure}
\begin{figure}[b!] % fig 10
\centerline{\hbox{\epsfig{file=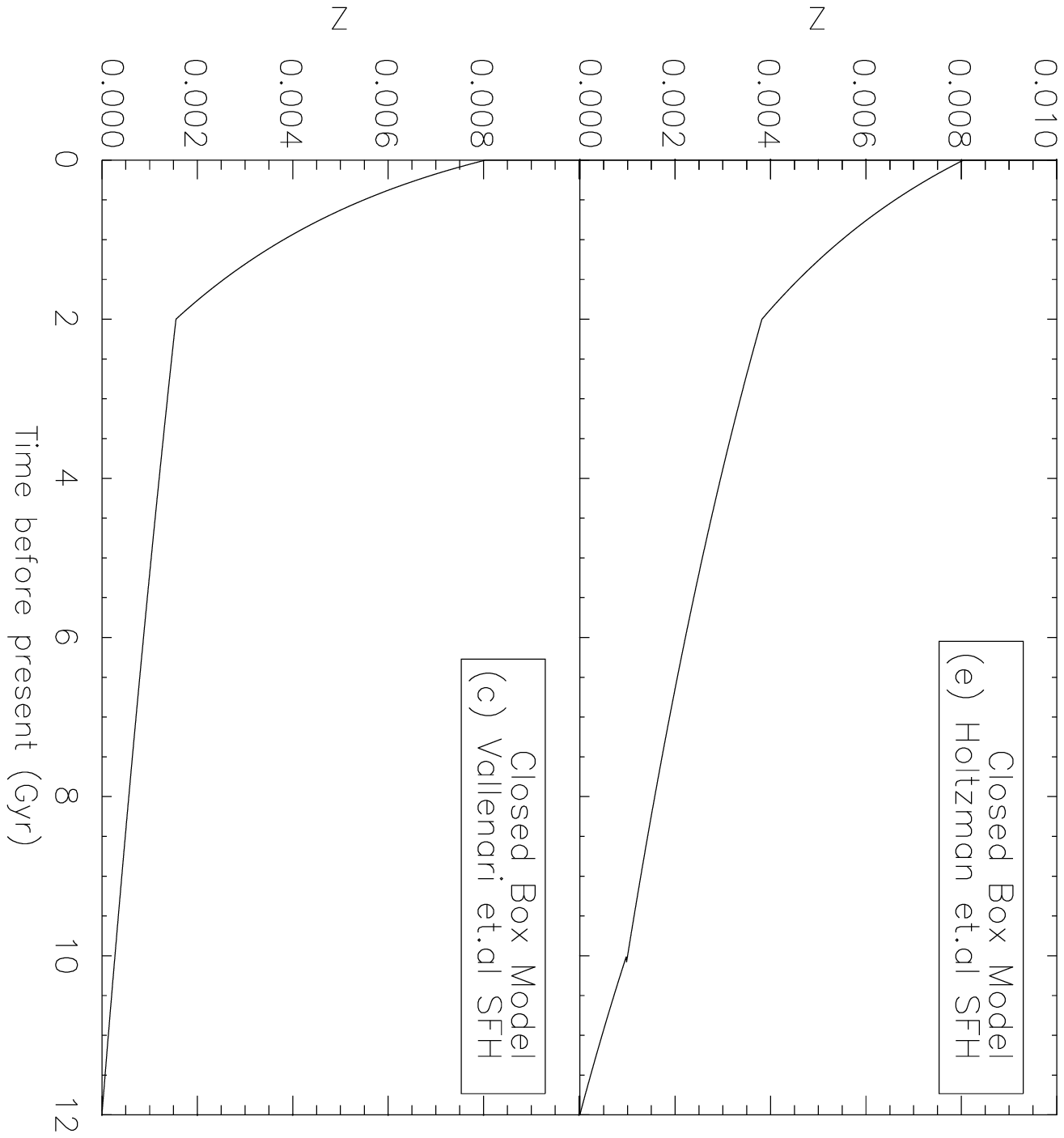, height=5.5 in, width=5.5 in }}}
\vskip 0.5 in
{Fig. 10.-}
\end{figure}
\begin{figure}[b!] % fig 11
\centerline{\hbox{\epsfig{file=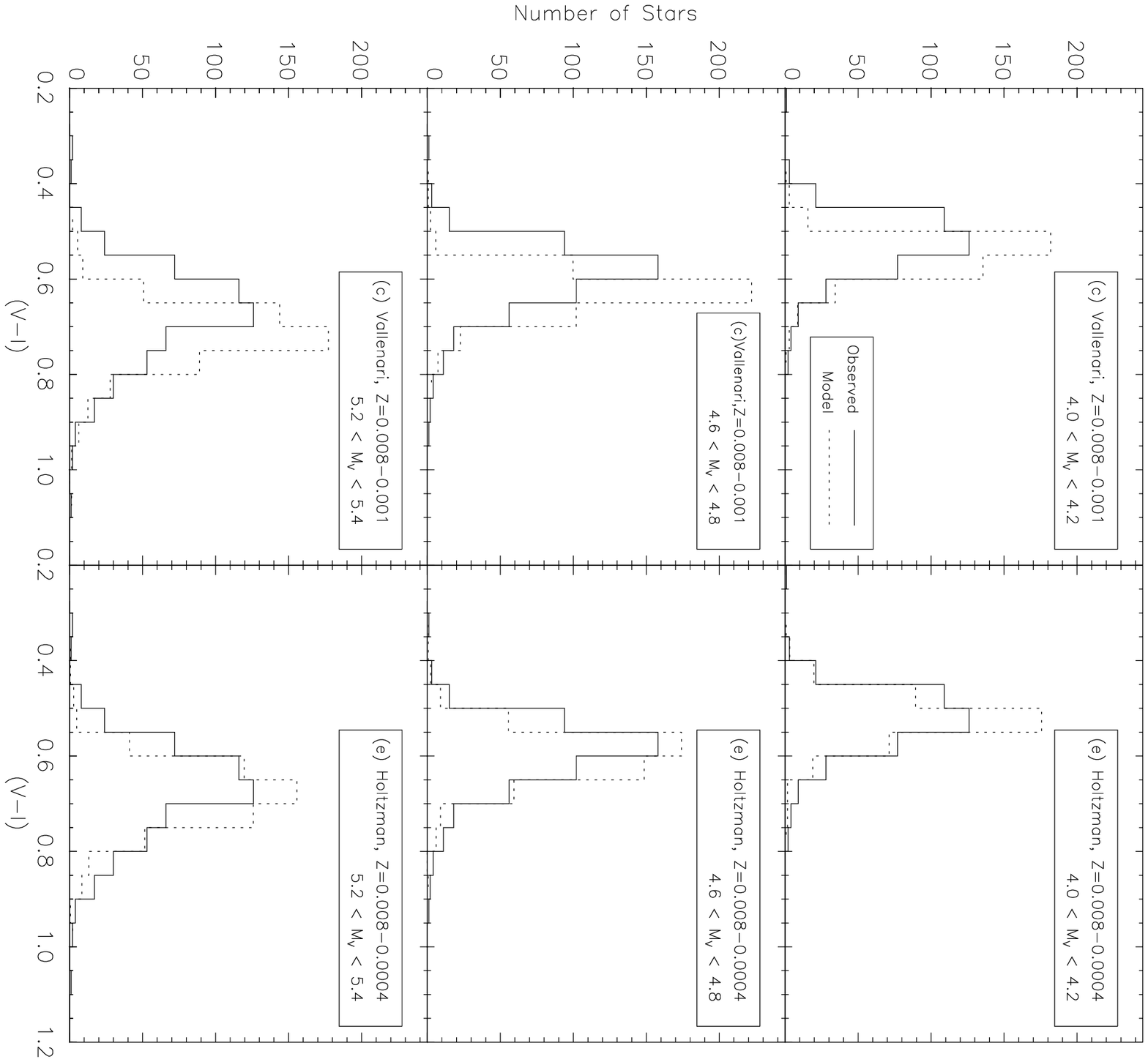, height=6 in, width=6 in }}}
\vskip 0.5 in
{Fig. 11.-}
\end{figure}


\newpage
\begin{references}
\reference{bbcf94} Bertelli, G., Bressan, A., Chiosi, C.,
Fagotto, F.
\& Nasi,
E. 1994, A\&AS, 106,275
\reference{bmcb92} Bertelli, G., Mateo, M., Chiosi, C., \&
Bressan,
A. 1992, ApJ, 388, 400
\reference{bt87} Binney, J. \& Tremaine, S. 1987, {\it {Galactic
Dynamics}}, (Princeton University Press: Princeton, N.J.)
\reference{cohe88} Cohen, R. S. \etal ApJ 331, L95 
\reference{dopi97} Dopita, M. \etal 1997, ApJ, 474, 188
\reference{egs97} Elson, R., Gilmore, G. \& Santiago, B. 1997,
MNRAS, 289, 157
\reference{gall96} Gallagher, J. S., \etal 1996, ApJ, 466, 732
\reference{gira95} Girardi, L. \etal 1995, A\&A, 298, 87
\reference{irw92} Iglesias, C. A., Rogers, F. J., \& Wilson, B.
G.
1992, ApJ, 397, 717
\reference{holt95a} Holtzman, J. A. \etal 1995a, PASP, 107, 156
\reference{holt95b} Holtzman, J. A. \etal 1995b, PASP, 107, 1065
\reference{holt97} Holtzman, J. A. \etal 1997, AJ, 113, 656
\reference{ll92} Luck, R. E. \& Lambert, D. L. 1992, ApJS, 79,
303
\reference{mf97} Madore, B. \& Freedman, W. 1997 ApJL, in press
\reference{olsz93} Olszewski, E.W. 1993, in ASP Conf. Proc. 48,
The Globular Cluster-Galaxy Connection, ed. G. Smith \& J. Brodie
(San Francisco:ASP), 351
\reference{pana91} Panagia \etal 1991, ApJ, 380, L23.
\reference{perr95} Perryman \etal 1995, A\&A, 304, 69.
\reference{reid91} Reid, N. 1991 AJ, 102, 1428.
\reference{sand61} Sandage, A. 1961, {\it {The Hubble Atlas of
Galaxies}}, (Carnegie Institution of Washington: Washington,
D.C.) 
\reference{sara95} Sarajedini, A., Lee, Y., Lee, D. 1995 ApJ
450, 712
\reference{si91} Schwering, P. G., \& Israel, F. P. 1991, A\&A,
246, 231
\reference{ss72} Searle, L., \& Sargent, W. L. W. 1972, ApJ, 173,
25
\reference{scs93} Salaris, M., Chieffi, A., \& Straniero, O.
1993, ApJ, 414, 580 
\reference{stap97} Stappers, B. J., \etal 1997, PASP, 109, 292
\reference{stry84} Stryker, L., 1984, ApJS, 55, 127
\reference{sunt97} Suntzeff, N. 1997, Bull. A. A. S., 190, 3501
\reference{vcbo96a} Vallenari, A., Chiosi, C. Bertelli, G., \&
Ortolani, S. 1996a, A\&A, 309, 358
\reference{vcba96b} Vallenari, A., Chiosi, C. Bertelli, G.,
Aparicio,
A., \& Ortolani, S. 1996b, A\&A, 309, 367
\reference{vand91} van den Bergh, S. 1991, ApJ, 369, 1
\reference{wll95} Westerlund, B. E., Linde, P. \& Lynga, G. 1995,
A\&A, 298, 39
\reference{zht97} Zaritsky, D. Harris, J., \& Thompson, I. 1997, Astro-Ph/9709055
\end{references}
\end{document}